\documentclass{aa}  
\usepackage{graphicx}
\usepackage{txfonts}
\usepackage{aalongtable}
\begin{document}
\title{Effective grain surface area in the formation of molecular hydrogen in interstellar clouds}

%\subtitle{Grain surface area in the formation of molecular hydrogen}

   \author{Sandip K. Chakrabarti\inst{1,2}, Ankan Das\inst{2}, K. Acharyya 
           \inst{2}, S. Chakrabarti \inst{2,3}
%          \and
%         C. Ptolemy\inst{2}\fnmsep\thanks{Just to show the usage
%         of the elements in the author field}
          }

   \offprints{Prof. Sandip K. Chakrabarti}

   \institute{ S. N. Bose National Centre for Basic Sciences, Salt Lake, 
              Kolkata 700098, India\\
             \email{chakraba@bose.res.in}
                   \and
                Centre for Space Physics, Chalantika 43, Garia Station Rd.,
               Kolkata, 700084, India\\
              \email{ankan@csp.res.in, acharyya@csp.res.in}
            \and
                Maharaja Manindra Chandra College, 20 Ramkanta Bose Street.
              Kolkata 700003, India\\
               \email{sonali@csp.res.in} 
             }

   \date{Received ; accepted }

% \abstract{}{}{}{}{} 
% 5 {} token are mandatory
  \abstract
{}
{In the interstellar clouds, molecular hydrogens are formed from atomic
hydrogen on grain surfaces. An atomic hydrogen hops around till it finds another
one with which it combines. This necessarily implies that the average 
recombination time, or equivalently, the effective grain surface area
depends on the  relative numbers of atomic hydrogen influx rate and the number of sites on the
grain. Our aim is to discover this dependency.}
{We perform a numerical simulation to study the recombination 
of hydrogen on grain surfaces in a variety of cloud conditions. 
We use a square lattice (with a periodic boundary condition) of various sizes on two 
types of grains, namely, amorphous carbon and olivine.}
{We find that the steady state results of our simulation match very well with those obtained from a
simpler analytical consideration provided the `effective' grain surface area is 
written as $\sim S^{\alpha}$, where, $S$ is the actual physical grain area and
$\alpha$ is a function of the flux of atomic hydrogen which is determined from our 
simulation. We carry out the simulation
for various astrophysically relevant accretion rates. For high accretion rates, small
grains tend to become partly saturated with $H$ and $H_2$ and the subsequent
accretion will be partly inhibited. For very low accretion rates, the number of sites to be
swept before a molecular hydrogen can form is too large compared to the actual number
of sites on the grain, implying that $\alpha$ is greater than unity.
}{}
%   {empty}
% % results heading (mandatory)
%  {empty}
% % conclusions heading (optional), leave it empty if necessary
%  {}


\titlerunning{Effective grain surface area}

\authorrunning{S. K. Chakrabarti et al. }

\maketitle
%
%________________________________________________________________

\section{Introduction}
It is long known that the formation of $H_2$ in the interstellar cloud takes 
place on the grain surfaces and are then released into the gas phase. Considerable 
studies were made since then to understand the real physical processes 
which are taking place both theoretically (e.g., Hollenbach, Werner \& Salpeter, 1971;
Takahashi, Matsuda \& Nagaoka, 1999; Biham et al. 2001) and
experimentally (e.g., Pirronello et al. 1999). A quantitative estimation
based on physisorption of the formation rate of $H_2$ was carried out by
Biham et al. (2001) and Green et al. (2001). It was found that a
significant production is possible in cooler ($\sim 10-25$K)
clouds (Rae et al. 2003). Cazaux \& Tielens (2002, 2004) use 
both physisorption and chemisorption, to demonstrate
that $H_2$ production is possible at high temperatures ($\sim 200-400$K) also.
Recently, Acharyya, Chakrabarti \& Chakrabarti (2005) and Acharyya \& Chakrabarti (2005)
computed the formation of $H_2$ from $H$ assuming various cloud conditions and using
rate and master equations presented by Biham et al. (2001).
If one considers only $H$ and $H_2$, then, {\it in a steady state},
the number of hydrogen atoms ($n_H$) and molecules ($n_{H2}$) on a grain surface
are obtained by equating the effective influx (or, accretion)
rate $\phi_H$ of $H$ with the rate at which they are used
up to form $H_2$ and/or get desorbed from the grain surface into the gas phase, i.e.,
$$
\phi_H= W_H n_H + 2 a_H n_H^2 ,
\eqno{(1a)}
$$
and
$$
W_{H2}n_{H2}= \mu a_H n_H^2 .
\eqno{(1b)}
$$
Here, $\phi_H$ is the effective flux (in units of number/s) of $H$ on the grain surface,
$W_H$ and $W_{H2}$ (both in units of s$^{-1}$) are the desorption co-efficients of  $H$
and $H_2$ respectively, $a_H$ (in units of s$^{-1}$) is the effective rate of recombination
of two $H$ atoms to form one $H_2$ molecule and $1-\mu$ is the fraction of
$H_2$ that are formed on the grain which are spontaneously released to the gas phase.
Solving these simple equations, we find in the steady state, number of $H$ 
and $H_2$ on a grain surface to be, 
$$
n_H=\frac{\sqrt{W_H^2 + 8 a_H\phi_H}-W_H}{4a_H},
\eqno{(2a)}
$$
$$
n_{H2}= \frac{\mu a_H n_H^2}{W_H}
\eqno{(2b)}
$$
We ignore $H_2$ formation by tunneling since the experiments indicate hopping to be the major process (see,
Pironello et al. 1999; Katz et al. 1999). In the literature, it is customary to define the 
effective rate of recombination to be,
$$
a_H = A_H/S,
\eqno{(3a)}
$$
(See, Acharyya, Chakrabarti, \& Chakrabarti, 2005  and references therein), where $A_H$ is the
hopping rate (inverse of the diffusion time $T_d =1/A_H$) on a grain surface and $S$ is the
number of sites on the grain surface (Biham et al. 2001). 
The argument for reducing the diffusion rate by a factor of $S$ is this: on an
average, there are $n=S^{1/2}$ number of sites in each direction of the
grain. Since the hopping is random, it would take a square of this, namely,
$n^2=S$ number of hopping to reach a distance located at $n=S^{1/2}$ sites away,
where, on an average, another $H$ is available. Thus, the effective recombination
rate was chosen to be $A_H/S$. However, when the flux is very high in a small grain,
many sites would be full and an $H$ need not hop a distance of $S^{1/2}$ to meet another one.
Similarly, when the accretion rate is too low, one $H$ may have to sweep the whole grain
several times before it can meet another $H$ to recombine. Thus, it is expected that the
effective rate of recombination would be a function of the number of sites as well as
the flux of $H$ on the grain. If we assume that the effective number of sites could be
written as $S^{\alpha_0}$, where $\alpha_0$ is a constant for a given flux and grain, then,
in principle, we could define the effective recombination rate to be,
$$
a_H=A_H/S^{\alpha_0},
\eqno{(3b)}
$$
and we expect that higher the accretion rate, lesser would be the value of $\alpha_0$. 
In the opposite limit, when the accretion rate is low, one 
would get $\alpha_0>1$ since the effective site number is higher 
than $S$. In order to see this effect we clearly need to actually 
compute time dependences of $H$ and $H_2$ by numerical simulation 
and derive $\alpha_0$ when a 'steady state' is reached. This we do in this paper.
                                                                           
In the next Section, we present the logic based on which the simulation is
carried out. In \S 3, we present our results
under various grain conditions and compare our results with the
expectation from the steady state (Eqs. 2ab). We show that these two results match
only if $\alpha_0$ is used in the steady state equations. Finally, in Section \S 4 we draw our conclusions.
                                                                           
\section{Numerical Simulation}
                                                                           
The simulation is based on several physical processes which might be taking place on
the grain surface. We assume each grain to be square in shape having
$S=n^2$ number of sites (where, $n$ is the number of sites in each
direction). We use the periodic boundary condition. We consider only the mono-layer
on the grain surface. Let $A_H = \nu\exp(-E_0/k_bT)$, the rate at which $H$ hops
from a site to the next, where, $\nu$ is the vibrational frequency,
$$
\nu= \sqrt{\frac{2 s E_d}{\pi^2 m_H}} \ s^{-1}.
$$
Here, $s \sim 10^{14}$ cm$^{-2}$ is the surface density of sites on a grain,
$m_H$ is the mass (in gm) of the $H$ atom and $E_d$ is the binding energy 
corresponding to physisorption (in meV). We choose the minimum time step to be $1/A_H$, the hopping time
and advance the global time by this time step. Let $F_H$ be the accretion rate of $H$
which increases the number of $H$ on a grain by sticking to it. 
Here, $F_H= A <V> N_H$ s$^{-1}$, $N_H$ is the number density (in cm$^{-3}$) of $H$ in the gas phase, 
$<V>$ (in cm$^{-2}$) is the average thermal velocity of the particle in the gas phase and $A$
is the physical grain surface area (in cm$^{2}$); when $10^2 \la N_H \la 10^7$,
and $V\sim 10^4$, we have $3\times 10^{-8} < F_H < 3 \times 10^{-3}$. We note that 
the rate at which the dropping of $H$ takes place on a grain is a very small
number. To handle this, we drop one $H$ after every $A_H/F_H$ number of hops. 
When $F_H$ is very low, we continued our simulation for a larger time interval till
a steady state is reached. Similarly, where $F_H$ is large enough so that $F_H/A_H >1$, we
drop several at each time step. In any case, for a given $F_H$, the effective 
accretion rate is given by $\phi_H=F_H (1-f_{grh}-f_{grh2})$, where $f_{grh}$ 
and $f_{grh2}$, are the fractions of the grain which are occupied by $H$ and $H_2$ respectively.
After $1/A_H$ seconds, $\phi_H/A_H$ number of hydrogen atoms are dropped on the grain.
The exact site at which one atom is dropped is obtained by a pair of random                
numbers ($R_x, R_y$; $R_x <1$ and $R_y <1$) obtained by a random number generator.
This pair would place the incoming hydrogen at ($i,j$)th grid, where, $i$ and $j$ 
are the nearest integers obtained using the {\it Int} function: $i=int(R_x*n+0.5)$ 
and $j=int(R_y*n+0.5)$. Each atom starts hopping with equal probability 
in all four directions. This direction is decided by another random number. When, during 
the hopping process, one atom enters a site which is already occupied by 
another atom, we assume that a molecule has formed and we increase 
the number of $H_2$ by unity and reduce the number of $H$ by two. However, 
when the atom enters a site occupied by an $H_2$, another random number is generated
to decide which one of the other nearest sites it is going to occupy.
Thermal evaporation of $H$ and $H_2$ from a grain surface are handled in the
following way. Let $W_H$ and $W_{H2}$ be the desorption co-efficients of hydrogen
and hydrogen molecule given by $\nu\exp(-E_1/k_bT)$ and $\nu\exp(-E_2/k_bT)$,
where $E_1$ and $E_2$ are the activation barrier energy (in meV) for desorption
of $H$ and $H_2$ molecule. Thus one atom is released to the
gas phase after every $1/W_H$ seconds and one molecule is being released
after every $1/W_{H2}$ seconds. We generate a random number $R_t$ for every $H$
present on the grain and release (at each time step, i.e., $1/A_H$ seconds) only
those for which $R_t< W_H/A_H$. A similar procedure is followed for the evaporation
of $H_2$ for which the criterion for evaporation was $R_t < W_{H2}/A_H$.
There could also be spontaneous desorption and a factor of 1-$\mu$ of $n_{H2}$ is
evaporated to the gas phase. Here too, a random number $R_s$ is
generated for each newly formed $H_2$ present on the grain. Those which satisfy
$R_s < 1-\mu$ are removed to the gas phase.
                                                                           
We continue our simulation for each type of grain and for each accretion rate
and temperature for a duration more than $10^6$s or more depending on the accretion rate
$F_H$, till a quasi-steady state is reached. After an initial transient
period, $n_H$ and $n_{H2}$, the number of $H$ and $H_2$ reach a quasi-steady
state with some fluctuations. These numbers can be used to compute the recombination
time. The values of the activation barrier energies $E_0$, $E_1$ and $E_2$ are
taken from Katz et al. (1999) and are given by, $E_0 = 24.7$ meV,
$E_1 = 32.1$ meV and $E_2 = 27.1$ meV for olivine and $E_0 = 44$ meV, $E_1 = 56.7$ meV 
and $E_2 = 46.7$ meV for amorphous carbon grains. For olivine,  $\mu = 0.33$ and for amorphous 
carbon $\mu = 0.413$ was used. 
                                                                           
\section{Recombination time scale as a function of grain and gas parameters}

\begin {figure}
\includegraphics[width=5.1cm]{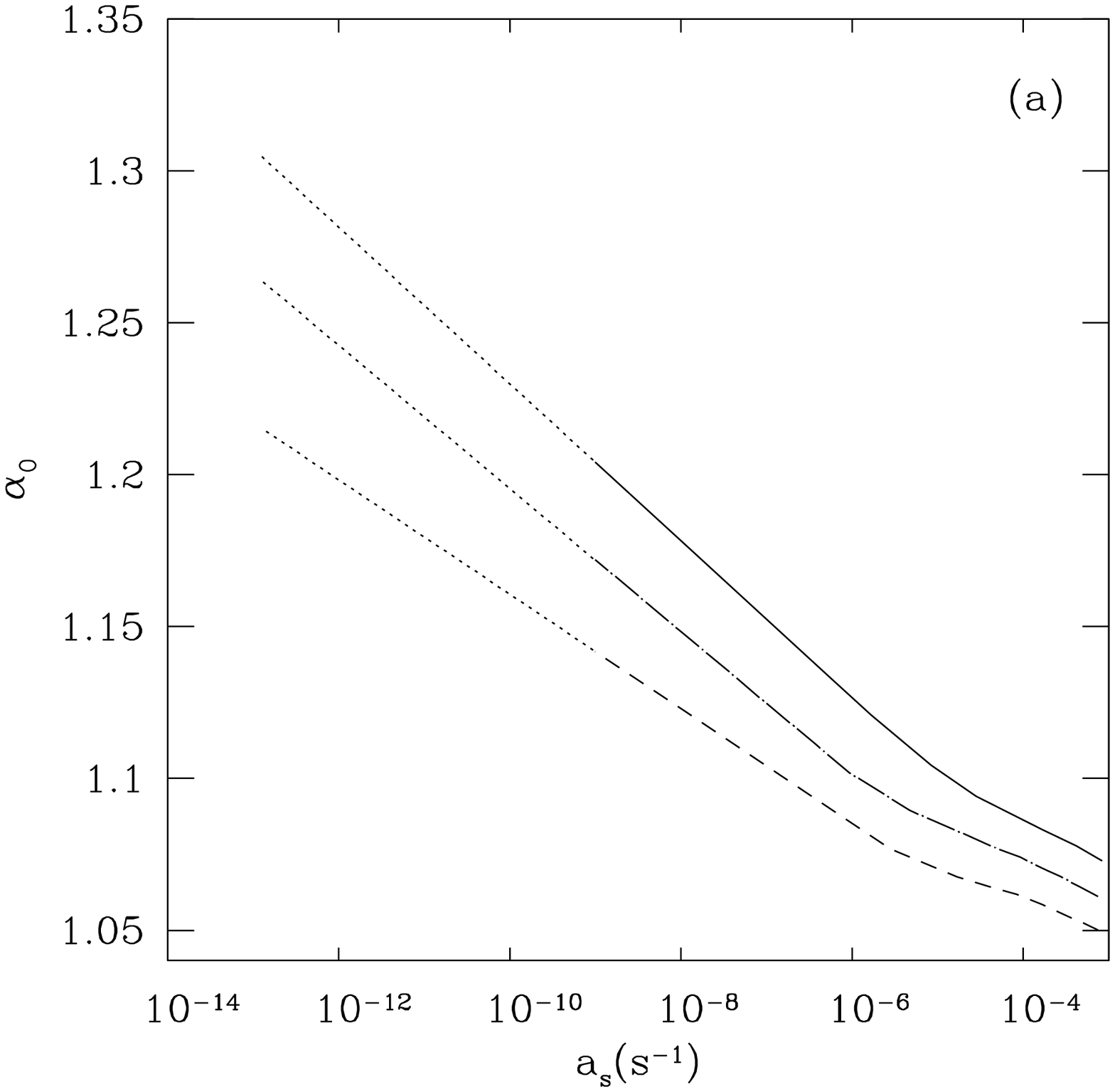}
\centering
\vskip -.6cm
\includegraphics[width=5.1cm]{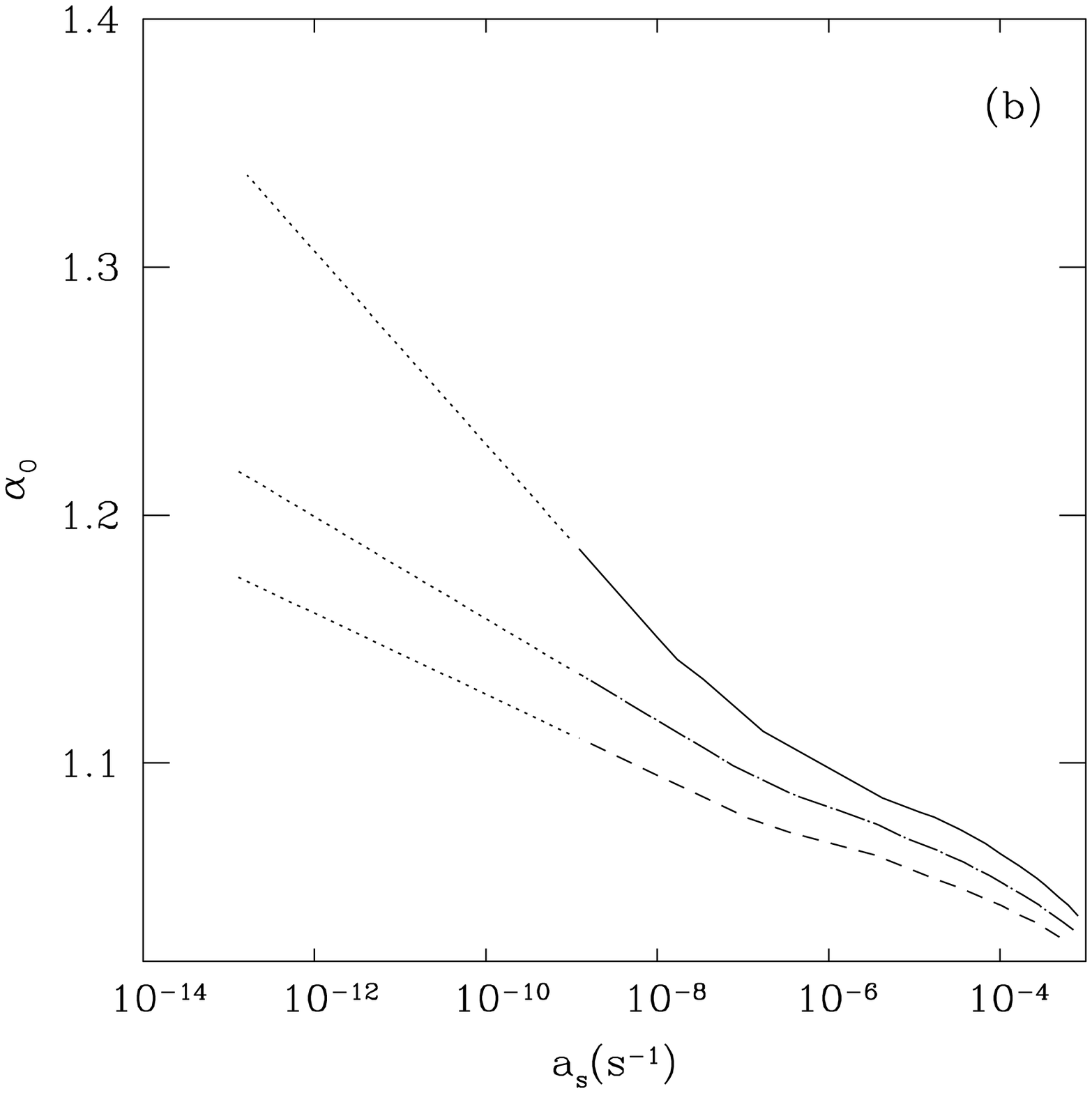}
\centering
\caption{{\bf (a-b)}: Variation of $\alpha_0$ as a function of the accretion rate per site
for an (a) amorphous carbon grain kept at $20$K and (b) olivine grain kept
at $10$K. The solid, dot-dashed and dashed curves are for $10^4$, $9 \times 10^4$
and $10^6$ sites respectively. As the accretion rate is reduced,
$\alpha_0$ starts to deviate from $1$ as the effective grain site number
goes up due to sweeping of the grain more than once before recombination takes place.
The deviation is highlighted using dotted curves by extrapolating at very
low accretion rates.}
\end{figure}
                                                                                                  
\begin {figure}
\includegraphics[width=5.1cm]{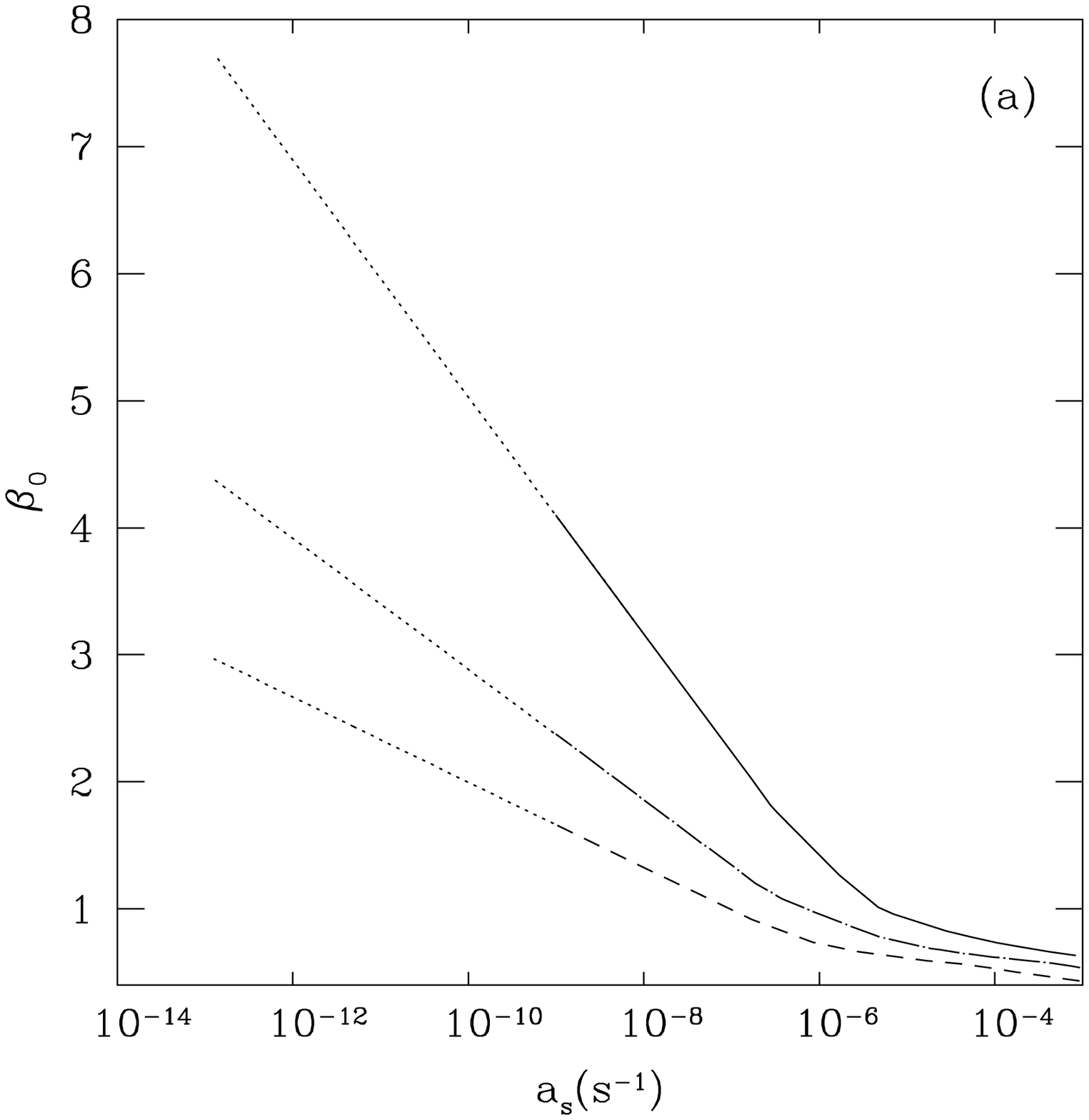}
\centering
\vskip -.6cm
\includegraphics[width=5.1cm]{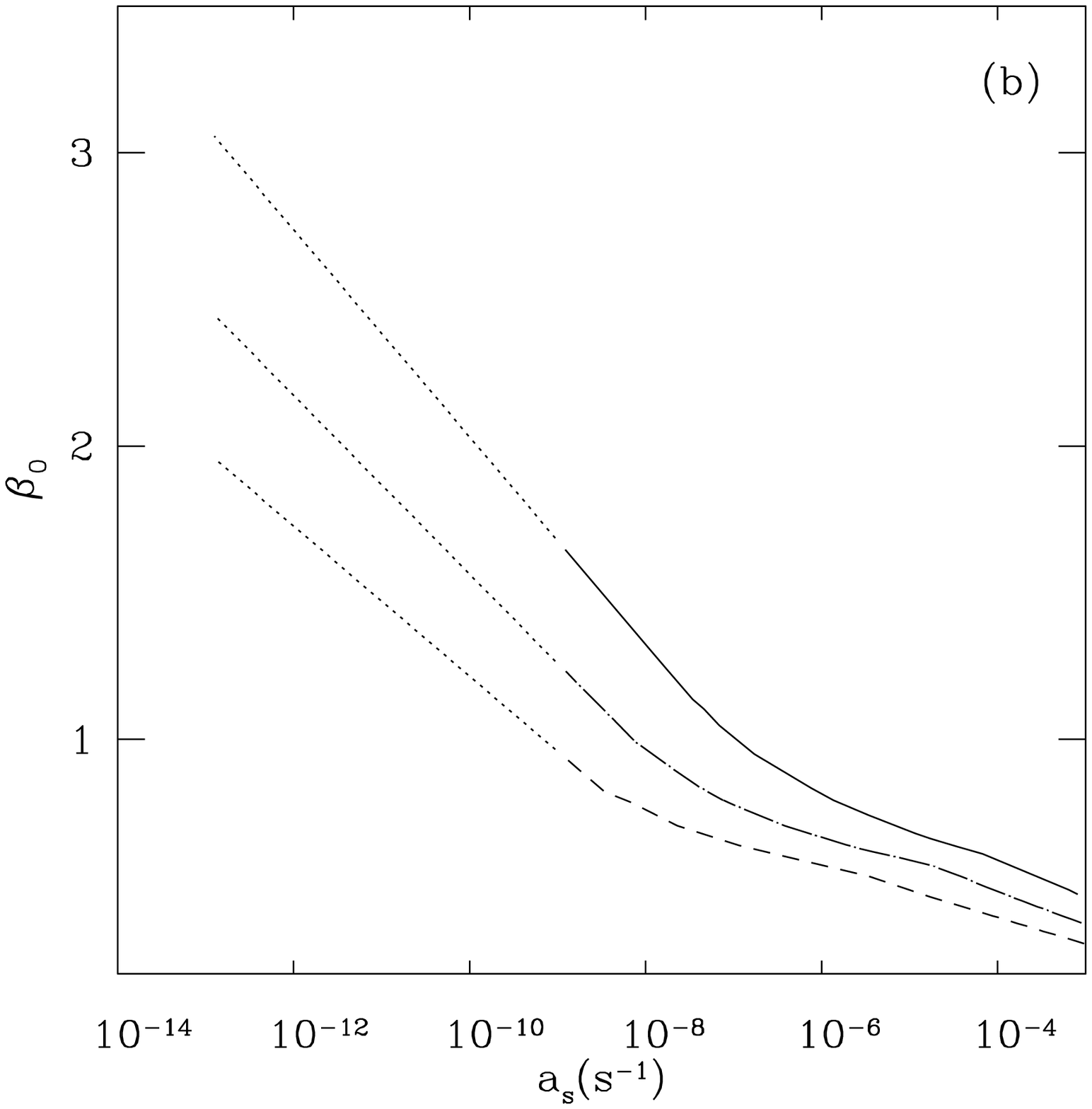}
\centering
\caption{{\bf (a-b)}: Variation of $\beta_0$ as a function of the accretion rate
per site for an (a) amorphous carbon grain kept at $20$K and (b) olivine grain
kept at $10$K. The solid, dot-dashed and dashed curves are for $10^4$,
$9 \times 10^4$ and $10^6$ sites respectively. As the accretion rate is reduced,
$\beta_0$ becomes very high compared to unity for the same reason as $\alpha_0$.
As the accretion rate is increased the grain becomes mostly occupied and an $H$
has to hop, on an average of $S^{1/2}$ times before an $H_2$ is produced.
Thus $\beta_0$ becomes almost $1/2$. The deviation is highlighted using dotted
curves by extrapolating at very low accretion rates.}
\end{figure}
                                                                                                  
\begin {figure}
\includegraphics[width=5.1cm]{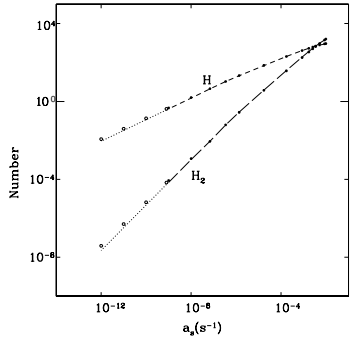}
\centering
\caption{: Comparison of the simulation results (dark circles) with
those obtained from analytical considerations [Eq. 2(a-b)] (dashed curves)
when suitable modification of the average recombination rate (Eq. 3b) is made. An olivine
grain of $10^4$ sites at a temperature of $10$K has been chosen in this comparison. Dotted
curves are drawn using analytical results for $\alpha_0$ extrapolated to very low accretion rates. }
\end{figure}
                                                                                           
Using the results of our simulation, we are now in a position to
compute a quantity $\alpha (t)$ defined by,
$$
\alpha (t) = log(\frac{2A_H n_H^2}{\phi_H-W_Hn_H})/log(S) .
\eqno{(4)}
$$
As $t\rightarrow \infty$, $\alpha(t)\rightarrow \alpha_0$.
If there were no effect of the accretion rate and the finite size of the grain,
then we would obtain $\alpha (t) = 1$ identically at all the time. However, we
find differently. In Figs. 1(a-b) we plot $\alpha_0$ for amorphous carbon
and olivine grains at $T=20$K and $T=10$K respectively as a function of the                                                    
accretion rate per site $a_s$. The solid, dot-dashed and dashed curves are
for $10^4$, $9 \times 10^4$ and $10^6$ sites respectively. For very very low
rates, it becomes impossible to carry out the simulations in a computationally finite time. 
We thus extrapolated the curves (dotted) for very low $a_s$ values. We find
that as the accretion rate per site goes down, $\alpha_0$ deviates from unity
significantly. Thus, clearly, for theoretical considerations,
the number of $H$ and $H_2$ should be calculated using Eq. (3b) and not Eq. (3a).
                                                                           
In this context, one important parameter $\beta$ may be defined as
the `catalytic capacity' of a grain which measures the efficiency of the
{\it formation} of $H_2$ on a grain surface for a given pair of $H$ residing on it.
Let $\delta N_{H2}$ be the number of $H_2$ {\it formed} in $\delta t$ time.
Since two hydrogen atoms are required to create one $H_2$,
the average rate of creation of one $H_2$ per pair of $H$ atom would be given by,
$$
<A_{H1}> = \frac{1}{2n_H} \frac{\delta N_{H_2}}{\delta t}.
\eqno{(5a)}
$$
We identify the inverse of this rate with the average formation rate given by,
$$
T_f(t) = S^{\beta(t)} /A_H.
\eqno{(5b)}
$$
Thus,
$$
S^\beta(t) = A_H/<A_{H1}>.
\eqno{(5c)}
$$
This yields $\beta(t)$ as a function of time as,
$$
\beta = log(A_H/<A_{H1}>)/log(S).
\eqno{(6)}
$$
Here too, we can define $\beta_0=\beta(t \rightarrow \infty)$.
Our goal would be to see if $\beta_0$ thus obtained actually varies with
grain parameters, accretion rate and temperatures.
We anticipated in Acharyya, Chakrabarti and Chakrabarti (2005) that
this exponent could be as low as $0.5$ or less. Indeed, this is what we see 
as well. In Figs. 2(a-b) we show $\beta_0$
as a function of the accretion rates per site for various grain
sites and for (a) amorphous carbon at $T=20$K and (b) olivine at $T=10$K grains 
respectively. The solid, dotted and dashed curves are for $10^4$, $9 \times 10^4$
and $10^6$ sites respectively as before. We find that $\beta_0$ can vary
anywhere between $\sim 1.5$ to $\sim 0.5$. Most interestingly, as the accretion
rate becomes high, the exponent becomes close to $0.5$, as if the recombination
is taking place on a one-dimensional grain! This is perhaps understandable, since as
the grain gets filled up, the maximum number of hopping in any direction should                                
not be more than $n$, i.e., the number of sites in any one direction of the grain.
                                                                           
It is interesting to compare the results of our simulation with those
obtained from the analytical considerations with and without our $\alpha_0$ factor in the
equation mentioned above. In Table 1, we present this comparison. We take an
olivine grain of $10^4$ sites at $10$K and vary the accretion rates. In Column 1,
we give the accretion rate per site of the grain. In Column 2 we present the
coefficient $\alpha_0$ which we derive from our simulation.  In Columns 3-5,
we present the number of $H$ as obtained by our simulation and the
modified equation (Eq. 3b) and the standard equations (Eq. 3a). Columns 6-8, we present
similar results for $H_2$. We find that our simulation matches more accurately with the analytical results
provided Eq. (3b) is chosen. If standard equation (Eq. 3a) is used, the deviation is very significant.

In Fig. 3, we compare the number of $H$ and $H_2$ on an olivine grain of $10^4$ sites
kept at $10$ K as a function of the accretion rate per site. The dashed curves
are drawn using Eqs. 2(a-b), and 3b (using $\alpha_0$ from the simulation)
while the dark circles are purely from our simulation. Using extrapolated
$\alpha_0$ (from Fig. 1b) we computed $n_H$ and $n_{H2}$ analytically
and see very good agreements with the simulation at lower rates also.  

\section {Concluding remarks}
                                                                           
In this paper, we carried out a numerical simulation to show that the
effective grain surface is not really the same as the physical grain surface.
The simulations have been carried out for amorphous carbon and olivine grains kept
at $20$K and $10$K respectively for a wide range of accretion rates.
In the literature, it is assumed that the recombination time is the
grain site number divided by the diffusion rate (Eq. 3a). However,
we find that this simplistic assumption is not valid especially when the
accretion rate of $H$ on the grains is very low.
In this case, an atomic hydrogen may have to sweep a grain several
times before meeting another atom to form a molecule thereby increasing the 
effective surface area. For very high rates,
the grain surfaces could be partially filled with $H$ and $H_2$ and search
for another $H$ need not take $S$ number of hops. We found that on an
average, $\alpha_0$, the index which determines effective number of sites,
could be $\ga 1$ for very low rates. We also find that $\alpha_0$ depends on the nature of the grains, 
the temperature and the grain site numbers. Thus the recombination time
is a complex function of these parameters. From the simulation results, we
defined another index $\beta$ which is a measure of the average
catalytic ability to form an $H_2$. This index is also found to be a strong function 
of the accretion rate. In fact, for very high rates
the index may go down to $0.5$ as anticipated in ACC05. 
A comparison of the recombination efficiency as obtained
from our procedure with that obtained from the standard consideration [Eq. 1(a-b)]
clearly shows a deviation easily attributable to the site/rate dependence
of $\alpha_0$ in a realistic case. The considerations based on which 
we came to our conclusion is so generic that we believe that similar dependence should
follow even for other grain surface reactions. This will be discussed elsewhere. We
believe that this important behaviour will have to be kept in mind
in future studies of chemical evolution in interstellar clouds in presence of grains.
                                                                           
Work of AD has been supported by a DST Project and that of KA is supported by an ISRO
Project with Centre for Space Physics.

{}

%\begin{longtable}{llllrrrr}
\begin{longtable}{|cc|ccc|ccc|}
\caption{\label{kstars} Comparison of $H$ and $H_2$ abundances in various methods}\\
\hline
Accretion Rate&  $\alpha_0$ &  & $H$ with &  &   & $H_2$ with  &  \\ \cline{3-5} \cline{6-8}
 per site  $A_s(S^{-1})$ & & Simulation &  $\alpha_0\ne 1$ & $\alpha_0=1$
& Simulation & $\alpha_0\ne 1$& $\alpha_0=1$\\
\hline
\endfirsthead
\caption{continued.}\\
\hline\hline
\endhead
\hline
\endfoot
$9.79 \times 10^{-3}$ & 0.9807& 935.75 & 903.30 &1008.83 & 1609.74  & 1581.09 &1650.50 \\
\hline
$8.58 \times 10^{-3}$ & 0.9857& 909.51 & 884.82&958.35  & 1442.57  & 1447.73  &1489.45 \\
\hline
$5.15 \times 10^{-3}$ & 1.0021& 791.82 & 784.50&775.71  & 940.92  & 978.55
 &975.84 \\
\hline
$3.43 \times 10^{-3}$ & 1.0127 & 694.71  & 692.13 &649.17 & 658.34  & 691.38   &683.45 \\
\hline
$2.58 \times 10^{-3}$ & 1.0191  & 627.84 & 626.46 &569.76  & 507.21  & 533.68    &526.46\\
\hline
$1.71 \times 10^{-3}$ & 1.0271  & 539.40 & 538.32 &471.95  & 349.42  & 366.19    &361.22\\
\hline
$8.58 \times 10^{-4}$ & 1.0385  & 408.56   & 407.07 &339.14 & 180.06  & 188.55    &186.52\\
\hline
$1.72 \times 10^{-4}$ & 1.0586  & 203.73  & 202.09 &154.06 & 37.40  & 38.62    &38.49\\
\hline
$1.72 \times 10^{-5}$ &  1.0783  &71.38  & 69.68 &48.75  & 3.72  & 3.83
&3.85\\
\hline
$1.37 \times 10^{-6}$ & 1.1013  & 21.41 & 21.10 &13.49  & 0.28  & 0.280
&0.2951\\
\hline
$1.72 \times 10^{-7}$ & 1.1170  & 7.32 & 7.20 &4.49 & 0.027  & 0.0290    &0.0327\\
\hline
$1.14 \times 10^{-7}$ & 1.1211  & 5.89 & 5.76 &3.59  & 0.018  & 0.018    &0.0209\\
\hline
$3.43 \times 10^{-8}$ & 1.1329  & 2.85 & 2.78  &1.79  & 0.0038  & 0.0037
&0.0052\\
\hline
\end{longtable}
\end{document}